# Topological insulator materials for advanced optoelectronic devices

Zengji Yue[1,2*], Xiaolin Wang[1,2] and Min Gu[3]

[1]*Institute for Superconducting and Electronic Materials, University of Wollongong, North Wollongong, New South Wales 2500, Australia*

[2]*ARC Centre for Future Low-Energy Electronics Technologies, Australia*

[3]*Laboratory of Artificial-Intelligence Nanophotonics, School of Science, RMIT University, Melbourne, Victoria 3001, Australia*

[*]Corresponding email: zengji@uow.edu.au

**Abstract**

Topological insulators are quantum materials that have an insulating bulk state and a topologically protected metallic surface state with spin and momentum helical locking and a Dirac-like band structure.(1-3) Unique and fascinating electronic properties, such as the quantum spin Hall effect, topological magnetoelectric effects, magnetic monopole image, and Majorana fermions, are expected from topological insulator materials.(4, 5) Thus topological insulator materials have great potential applications in spintronics and quantum information processing, as well as magnetoelectric devices with higher efficiency.(6, 7) Three-dimensional (3D) topological insulators are associated with gapless surface states, and two-dimensional (2D) topological insulators with gapless edge states.(8) The topological surface (edge) states have been mainly investigated by first-principle theoretical calculation, electronic transport, angle-resolved photoemission spectroscopy (ARPES), and scanning tunneling microscopy (STM).(9) A variety of compounds have been identified as 2D or 3D topological insulators, including HgTe/CdTe, $Bi_2Se_3$, $Bi_2Te_3$, $Sb_2Te_3$, $SmB_6$, BiTeCl, $Bi_{1.5}Sb_{0.5}Te_{1.8}Se_{1.2}$, and so on.(9-12)

On the other hand, topological insulator materials also exhibit a number of excellent optical properties, including Kerr and Faraday rotation, ultrahigh bulk refractive index, near-infrared frequency transparency, unusual electromagnetic scattering and ultra-broadband surface plasmon resonances. In details, Dirac plasmon excitations have been observed in $Bi_2Se_3$ micro-ribbon arrays at the THz frequency.(13) Ultraviolet and visible frequency plasmonics have been observed in nanoslit and nanocone arrays of $Bi_{1.5}Sb_{0.5}Te_{1.8}Se_{1.2}$ crystals.(14, 15) High transparency has been observed in nanometer scale $Bi_2Se_3$ nanoplates. Ultrahigh refractive index has been observed in the bulk of $Bi_{1.5}Sb_{0.5}Te_{1.8}Se_{1.2}$ crystals and $Sb_2Te_3$ thin films.(15, 16) These excellent optical properties enable topological insulator materials being capable of designing various optoelectronic devices, including plasmonic solar cells, ultrathin holograms, plasmonic and Fresnel lens, broadband photodetectors, and nanoscale waveguides. In this chapter, we focus on the excellent electronic and optical properties of topological insulator materials and their wide applications in advanced optoelectronic devices.(17)

## 1. Excellent electronic properties

### 1.1 Quantum spin Hall effect

The quantum spin Hall effect in 2D topological insulators was firstly proposed by Kane and Zhang.(18, 19) Then, the novel effect was predicted in mercury telluride–cadmium telluride (HgTe/CdTe) semiconductor quantum wells by König *et al.* in 2006.(4) The HgTe/CdTe quantum wells have well-known strong spin-orbital coupling. When the thickness of the HgTe layer is smaller than 6.3 nm, the 2D electronic states have the normal band order, but when the thickness is larger than 6.3 nm, the 2D bands have an inversion with a quantum phase transition between the trivial insulator and the quantum spin Hall insulator.

Molenkamp *et al.* experimentally realized the quantum spin Hall effect in HgTe quantum wells in 2007.(20) In the quantum well, two edge states with opposite spin polarization counter propagate at opposite edges. And, the quantum well also has non-local edge channel transport in the quantum spin Hall regime at zero external magnetic field.(21) These measurements confirm that the quantum transport through the helical edge channels is non-dissipative. And the topological protection of the edge states can't be destroyed by weak time reversal symmetric perturbations. The electrons in edge states are the absence of elastic backscattering and robustness against disorder for surface transport.

### 1.2 Topological magnetoelectric effects

Topological magnetoelectric effect is the phenomenon of magnetic polarization induced by applying an external electric field, or electric polarization induced by applying an external magnetic field. To obtain the topological magnetoelectric effect, a time-reversal-symmetry-breaking gap for the side surface is necessary.(22) A ferromagnetic layer with magnetization pointing out of the cylinder's surface induces a gap on the surface of the topological insulator, which has a fixed Hall conductance, $\sigma_H = (n +1/2) e^2/h$. When an electric field is applied parallel to the cylinder, a circulating current $j$ can be induced on the interface. This current is identical to the current generated by a constant magnetization $M$ which is anti-parallel to the electric field $E$. On the contrary, when a magnetic field $B$ is applied parallel to the cylinder, a circulating current is produced parallel to the interface, which induces a Hall current $j$ parallel or anti-parallel to the magnetic field *B*. As a result, charge density is accumulated on the top and bottom surfaces and induces the charge polarization.(2)

## 1.3 Magnetic monopole image

The topological magnetoelectric effect can be used to generate magnetic monopole image through putting an electric charge near a topological surface state.(23) When an electric charge is put on the surface of a 3D topological insulator with gapped surface states by time-reversal-symmetry breaking, the electric charge will polarize the bulk dielectric, and an image electric charge appears inside the topological insulator. An image magnetic monopole will also appear inside the topological insulator.(23)

The magnetic field generated by the image magnetic monopole has been experimentally measured and can be observed by a magnetic force microscope. A scanning magnetic force microscope tip can be applied to detect the magnetic field distribution of the image monopole. The magnetic field contribution from the magnetic monopole can also dominate and be distinguished from the contribution from surface impurities and roughness.

## 1.4 Topological superconductors

When topological insulators connect with ordinary superconductors, topological superconductors appear due to the correlated interface states and the proximity effect.(24) Such topological superconductors are predicted to host Majorana fermion excitations.(1) A Majorana fermion is a fermion that is its own antiparticle and was first predicted in the 1930s. The zero energy Majorana bound state is the simplest non-Abelian excitation and is associated with a vortex in a spinless superconductor. A Majorana zero mode has been proposed that can be realized in a superconducting vortex core by making use of the surface states of 3D topological insulators.(25) Topological protection and non-Abelian exchange statistics make the Majorana fermions promising for quantum computing. Signatures of Majorana fermions have been reported in quantum wires coupled to conventional superconductors.(26-28) Many experiments have been conducted to observe the elusive Majorana states.(27, 29)

Recently, chiral Majorana fermion modes was observed in a quantum anomalous Hall insulator–superconductor structure.(30) He *et al*. demonstrated the existence of one-dimensional chiral Majorana fermion modes in the hybrid system of a quantum anomalous Hall insulator thin film coupled with a superconductor. They conducted the transport measurements and found half-integer quantized conductance plateaus at the locations of magnetization reversals. The transport signature provided a strong evidence of the Majorana fermion modes. The discovery of Majorana fermion could pave a way for producing future topological quantum computing.

## 1.5 Quantum Anomalous Hall effects

The topological surface states of 3D topological insulators are protected by time-reversal symmetry and are robust against non-magnetic disorder. Nevertheless, surface states open up a gap in the presence of time-reversal symmetry breaking perturbations, and the Dirac electrons become massive. Magnetic impurities such as Fe, Co, and Mn dopants will induce a surface state gap. Moreover, theoretically, the quantized anomalous Hall effect can emerge in magnetic topological insulators.(31)

Such a quantum anomalous Hall effect was observed experimentally soon after it was theoretically predicted in thin films of chromium-doped $(Bi,Sb)_2Te_3$.(32) The gate-tuned anomalous Hall resistance reaches the predicted quantized value of $h/e^2$ at zero magnetic field. Under a strong magnetic field, the longitudinal resistance vanishes whereas the Hall resistance remains at the quantized value. The realization of the quantum anomalous Hall effect could lead to the development of next generation low-power-consumption electronics.

## 1.6 Giant magnetoresistance effects

Electronic transports are significant not only for the fundamental understanding of electronic properties of materials but also for promoting their applications in practical electronic and optoelectronic devices.(33, 34) Giant magnetoresistance was reported in a variety of 3D topological insulators.(35-41) The Dirac fermions of surface states suggest enhanced quantum corrections of magnetoconductance.(42) Spin-momentum locked surface states always show weak antilocalization effect effects due to strong spin-orbit coupling. Aharonov–Bohm oscillations have also been observed in $Bi_2Se_3$ and $Bi_{1-x}Sb_x$ topological insulators.(43, 44) Non-saturating positive linear magnetoresistance at high fields was observed in $Bi_2Te_3$ films and $Bi_2Se_3$ nanoribbons.(45-47) Room temperature giant and linear MR were observed in the topological insulator $Bi_2Te_3$ in the form of nanosheets with a few quintuple layers.(48) The giant, linear magnetoresistance that was achieved was as high as over 600% at room temperature, without any sign of saturation at measured fields up to 13 T. The observed linear magnetoresistance was attributed to the quantum linear magnetoresistance model developed by Abrikosov.(49, 50)

 Actually, the transport behavior of topological insulators reflects the combined contributions of surface states and bulk states due to the metallic bulk. A weak localization effect emerges when the time reversal symmetry is broken and a gap opens in surface states.(51-53) Competition between weak antilocalization effect and weak localization effect has been observed in the topological insulator thin film $Bi_{2-x}Cr_xSe_3$ and $(Bi_{0.57}Sb_{0.43})_2Te_3$.(54-56) The weak localization effect has been attributed to the 2D quantized channels of bulk states.(56)

## 1.7 Shubnikov-de Haas (SdH) effects

The surface states of 3D topological insulators have high carrier mobility and the Shubnikov-de Haas effect can appear in the presence of very intense magnetic fields.(57) The Shubnikov-de Haas effect is a macroscopic manifestation of the inherent quantum mechanical nature of matter. The Shubnikov-de Haas effect is a convincing tool for characterizing quantum transport in electronic materials.(58) It can be used to isolate the surface carriers and to determine their mobility and effective mass.(59) Shubnikov-de Haas effects were first observed in 3D topological insulator $Bi_2Te_3$ single crystals.(43) The surface mobility of up to 9000 to 10,000 $cm^2/V \cdot s$ was obtained based on Shubnikov-de Haas quantum oscillations, which is substantially higher than in the bulk. The obtained Fermi velocity of $4 \times 10^5$ m/s agrees with the results from angle-resolved photoemission measurements. Shubnikov-de Haas quantum oscillations have also been observed in 3D topological insulators, $Bi_2Te_2Se$ crystals, $Bi_2Te_3$ nanowires, and YPdBi crystals.(57, 60, 61)

## 2. Excellent optical properties

## 2.1 Ultrahigh bulk refractive index

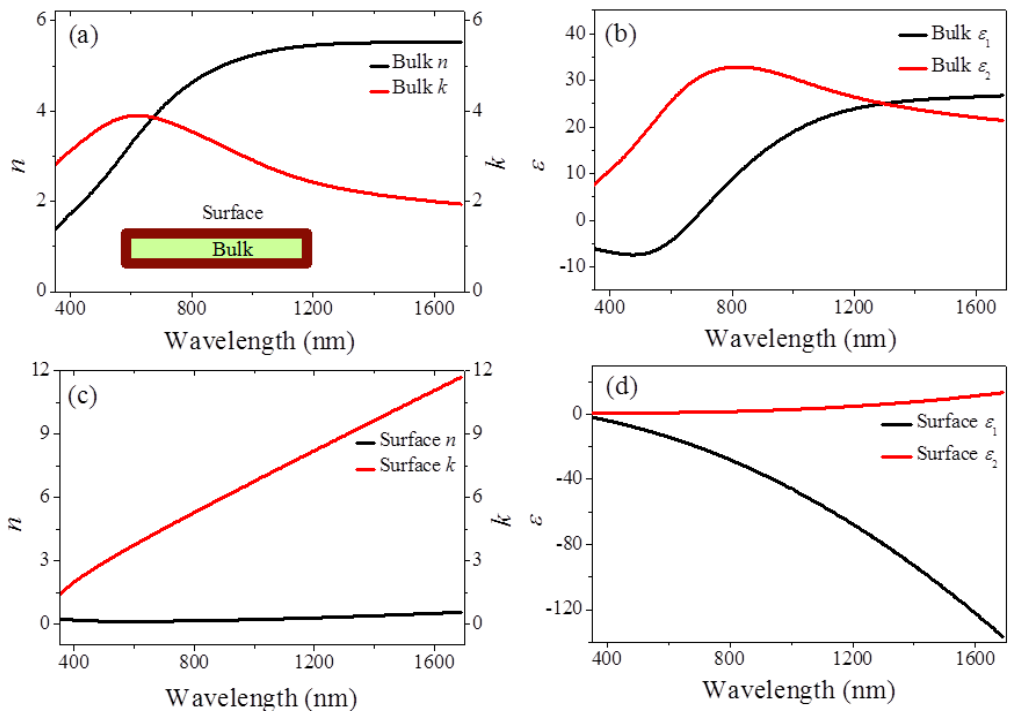

Figure 1. Optical parameters of $Bi_{1.5}Sb_{0.5}Te_{1.8}Se_{1.2}$ single crystals. (a-b) Refractive index *n*, extinction coefficient *k*, and dielectric function *ε* of the insulating bulk of $Bi_{1.5}Sb_{0.5}Te_{1.8}Se_{1.2}$ crystals. (c-d) Refractive index *n*, extinction coefficient *k*, and dielectric function *ε* of the metallic surface of $Bi_{1.5}Sb_{0.5}Te_{1.8}Se_{1.2}$ crystals. Reprinted with permission from [Yue, 2016], Science Advances 2 (2016) e1501536. © 2017, AAAS.

Optical constants are the basic optical parameters of materials that define the interaction of incident light and materials. Yue, *et al.* measured the refractive index *n* and extinction coefficient *k* of cleaved flat $Bi_{1.5}Sb_{0.5}Te_{1.8}Se_{1.2}$ crystal sheets by using a spectroscopic ellipsometer. They discovered that $Bi_{1.5}Sb_{0.5}Te_{1.8}Se_{1.2}$ crystal holds low refractive index in the surface but ultrahigh refractive index in the bulk in near infrared frequency.(15) This excellent optical property makes topological insulator materials promising for designing novel near-infrared optoelectronic devices.

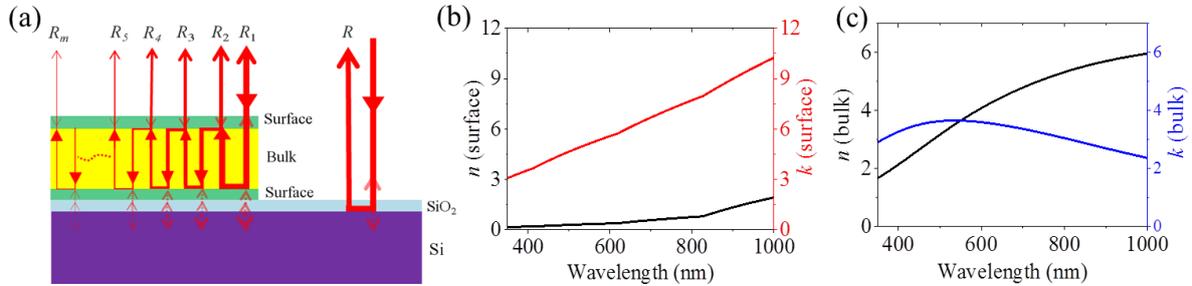

Figure 2. Physical mechanism of the $Sb_2Te_3$ thin film cavity. Diagram of internal light multiple reflections in the resonant cavity of the $Sb_2Te_3$ thin film. The refractive index *n* and extinction coefficient *k* of the surface and bulk of $Sb_2Te_3$ thin film. Reprinted with permission from [Yue, 2017], Nature Communications, (2017) 15354. © 2017 Macmillan Publishers Limited, part of Springer Nature.

In additions, Yue, *et al.* also discovered that the unequal refractive index in surface and bulk of topological insulator thin films could generate intrinsic resonant cavity.(16, 62, 63) The multilayer structure of the $Sb_2Te_3$ thin film on a Si substrate is schematically shown in Figure 2(a). The dielectric bulk of the $Sb_2Te_3$ thin film is sandwiched within the two metallic surface layers. The refractive index *n* and extinction coefficient *k* of the surface layers and the bulk are unequal due to different electronic property.(63) With the unequal refractive index, the $Sb_2Te_3$ thin film acts as an intrinsic optical resonant cavity.(64) Two surface layers serve as two reflectors. The bulk behaves as an optical resonant cavity. Thus an incident light beam can be reflected multiple times between two surface layers and partially be confined in the bulk. The phase modulation of the reflected light beam from the resonant cavity can be enhanced.

## 2.2 Near-infrared transparency

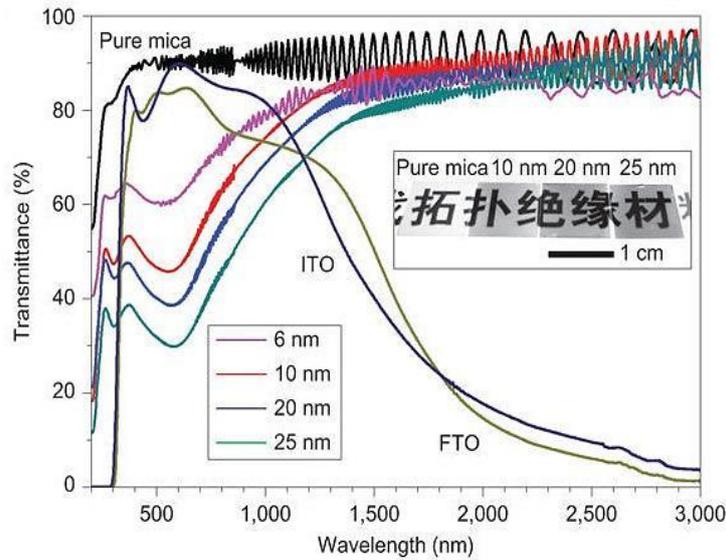

Figure 3. Spectroscopy characterization of the $Bi_2Se_3$ nanosheets on thin mica sheet substrates. Ultraviolet–visible-infrared spectra of $Bi_2Se_3$ nanosheets with different thickness, mica substrate, indium tin oxide (ITO) and fluorine tin oxide (FTO). The obvious oscillatory in the transmission spectra may result from Fabry–Perot interference effects. Reprinted with permission from [Yue, 2017].(65).

Peng, *et al.* demonstrated near-infrared transparent flexible electrodes based on few-layer $Bi_2Se_3$ nanostructures on mica. They found that the $Bi_2Se_3$ nanosheets exhibit a transparency of more than 70% over a wide range of wavelengths. Furthermore, the $Bi_2Se_3$ nanosheets were used for as transparent electrodes. The electrodes show high chemical and thermal stabilities as well as excellent mechanical durability. These features make the $Bi_2Se_3$ nanosheets promising candidates for novel optoelectronic devices. (65)

## 2.3 Faraday rotation and unusual electromagnetic scattering

Faraday rotation was predicted in topological insulator surface.(22) It results from topological magnetoelectric effects in magnetic topological insulators. In principle, it is possible to find a topological insulator with a larger gap which can support an accurate measurement of Faraday rotation. Similar proposals as above can also be worked out for the rotation of reflected wave Kerr effect. In additions, replace the ferromagnetic layers by paramagnetic materials with large susceptibilities and apply an external magnetic field to polarize them. In this case the magnetization is proportional to magnetic field, such that the Faraday rotation contributed by the bulk is also proportional to magnetic field. Unusual electromagnetic scattering was predicted from topological insulator nanoparticles.(66)

## 2.4 Ultra-broadband plasmon excitations

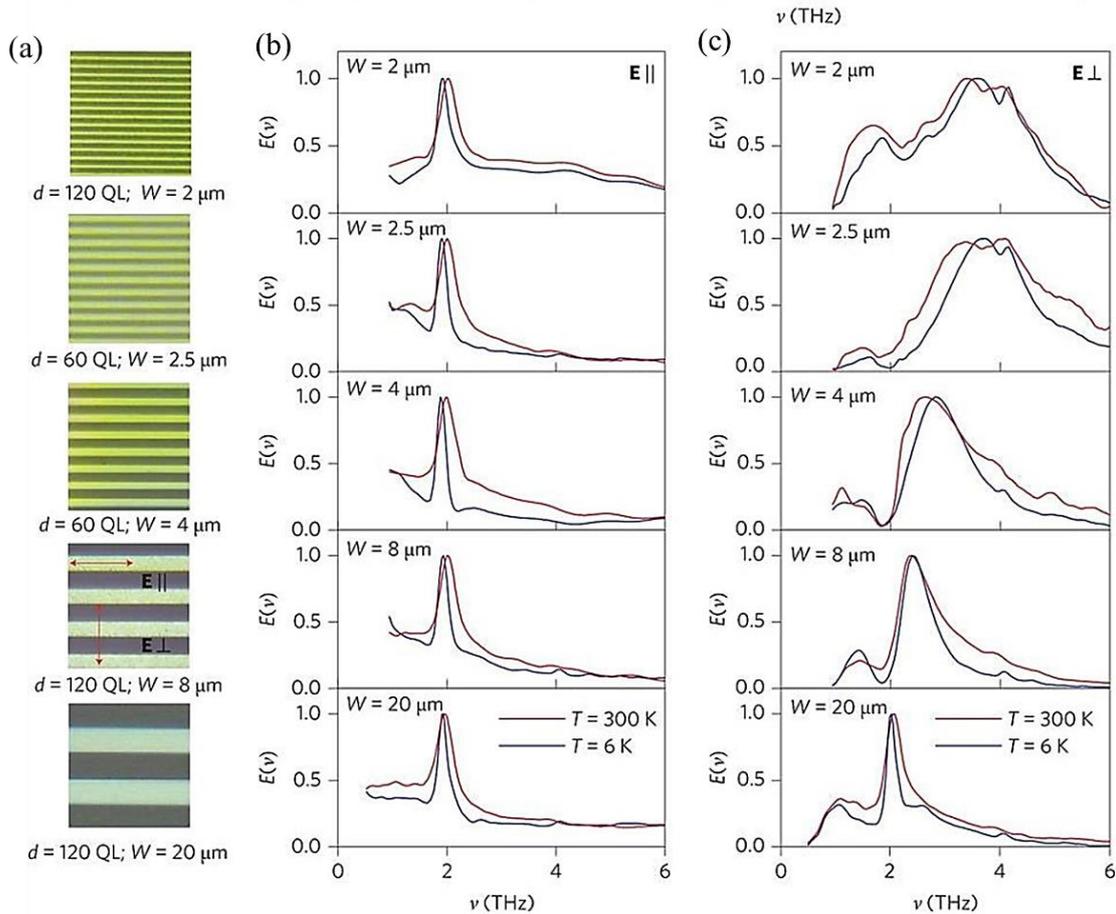

Figure 4. Extinction coefficients of the microribbon arrays of topological insulator $Bi_2Se_3$ in the terahertz range. (a) Optical microscope image of the 5 patterned films with different width and period. (b-c) Extinction coefficient of the 5 patterned films with the radiation electric field applied parallel and perpendicularly to the ribbons. Reprinted with permission from [Yue, 2017], (13).

Plasmons are quantized collective oscillations of electrons and have been mainly observed and investigated in noble metals. Plasmons have been widely applied in various optical devices from ultraviolet to THz frequency.(67-69) Dirac plasmons from massless electrons are promising for novel tunable plasmonic devices.(70) They exist in 2D materials like grapheme and semiconductors. Dirac plasmon excitations have also been observed in binary $Bi_2Se_3$ at the THz frequency.(13) The $Bi_2Se_3$ was prepared in thin micro-ribbon arrays of different widths and periods. The linewidth of the plasmon was found to remain nearly constant at temperatures between 6 K and 300 K.

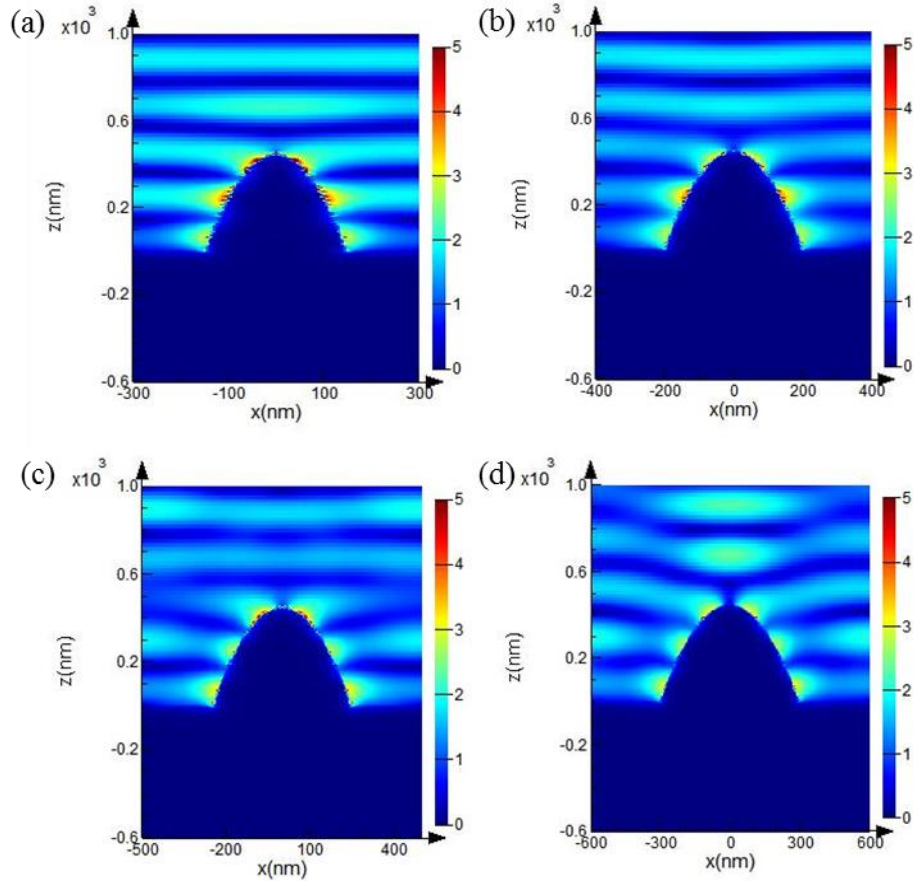

Figure 5. Simulation of electromagnetic field distribution in $Bi_{1.5}Sb_{0.5}Te_{1.8}Se_{1.2}$ nanocone arrays using the Finite-difference time-domain (FDTD) method. The plasmon resonances are localized and enhanced on the surfaces of $Bi_{1.5}Sb_{0.5}Te_{1.8}Se_{1.2}$ nanocones. Nanocone arrays have variable diameter and period. Reprinted with permission from [Yue, 2016], Science Advances 2 (2016) e1501536. © 2017, AAAS.

Ultraviolet and visible frequency plasmonics have been observed in nanoslit arrays and grating of bulk-insulating $Bi_{1.5}Sb_{0.5}Te_{1.8}Se_{1.2}$ crystals.(14, 15) Yue, *et al.* reported a conic nanostructure made of $Bi_{1.5}Sb_{0.5}Te_{1.8}Se_{1.2}$ crystals. They showed that the insulating bulk had an ultrahigh refractive index of up to 5.5 in the near-infrared frequency range. The metallic surface presented plasmonic excitations and strong backward light scattering in visible frequency range. Through integrating the nanocone arrays into a-Si thin film solar cells, up to 15% enhancement of light absorption was obtained.

Zhao, *et al.* have studied actively tunable visible surface plasmons in $Bi_2Te_3$ nanoplates electron energy-loss spectroscopy and cathodoluminescence spectroscopy.(71, 72) The observed plasmons in the visible range were mainly from metallic surface states of $Bi_2Te_3$. Infrared Yuan, et al. have investigated infrared nanoimaging of surface metallic plasmons in of $Bi_2Te_3$ nanoplates using scattering-type scanning near-field optical microscopy.(73) They discovered near-field patterns of bright outside fringes also originated from the surface-metallic plasmonic behavior at mid-infrared frequency.

With the metallic surface and insulating bulk, topological insulator materials provide an excellent platform for the realization of a new type of nanostructures that could combine the fascinating properties of plasmonic metallic nanostructures and dielectric nanostructures. With these features, the plasmonic topological insulator nanostructures pave a way for designing low-loss and high-performance optical devices, like visible to infrared detectors or sensors.(74)

## 2.5 Polarized light induced photocurrent

McIver, *et al.* reported the control of topological insulator photocurrents with light.(75) They showed that illuminating the $Bi_2Se_3$ with circularly polarized light generates a photocurrent that originates from topological helical Dirac fermions, and that reversing the helicity of the light reverses the direction of the photocurrent. They also observe a photocurrent that is controlled by the linear polarization of light and argue that it may also have a topological surface state origin. This approach may allow the probing of dynamic properties of topological insulators and lead to novel opto-spintronic devices.

## 2.6 Broadband optical nonlinear response

Chen, *et al.* studied the nonlinear response of $Bi_2Te_3$ at both the optical and microwave band. They demonstrated optical saturable absorption property of $Bi_2Te_3$ from 800 nm to 1550 nm. $Bi_2Te_3$ shows a saturation intensity of ~12 $\mu W/cm^2$ and a normalized modulation depth of ~70%. They argued that the optical saturable absorption in TI is a natural consequence of the Pauli-blocking principle of the electrons filled in the bulk insulating state.(76)

Giorgianni, *et al.* demonstrates an electromagnetic-induced transparency in $Bi_2Se_3$ under the application of a strong THz electric field. This effect, concomitantly determined by harmonic generation and charge-mobility reduction, is exclusively related to the presence of Dirac electron at the surface of $Bi_2Se_3$, and opens the road towards tunable THz nonlinear optical devices based on TI materials.(77)

The third order nonlinear optical property of $Bi_2Se_3$ was investigated under femto-second laser excitation.(78) When excited at 800 nm, the TI $Bi_2Se_3$ exhibits saturable absorption with a saturation intensity of 10.12 $GW/cm^2$ and a modulation depth of 61.2%, and a giant nonlinear refractive index of $10^{-14}$ $m^2/W$. This work suggested that the $Bi_2Se_3$ is a promising nonlinear optical material and can find potential applications from passive laser mode locker to optical Kerr effect based photonic devices.

## 3. Advanced optoelectronic devices

### 3.1 Plasmonic solar cells

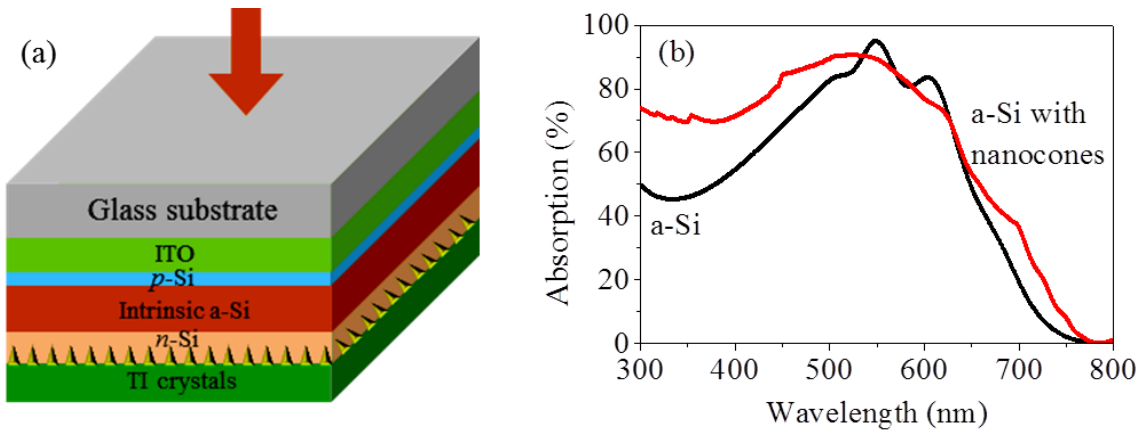

Figure 6. Plasmon resonances enhanced light absorption in ultrathin a-Si solar cells simulated using FDTD. (a) Structure of ultrathin a-Si solar cells with $Bi_{1.5}Sb_{0.5}Te_{1.8}Se_{1.2}$ nanocone arrays. The arrays achieve broadband enhancements of light absorptions in the visible frequency range. Reprinted with permission from [Yue, 2016], Science Advances 2 (2016) e1501536. © 2017, AAAS.

### 3.2 Nanometric holograms

Yue, *et al.* discovered that nanometric topological insulator thin films act as an intrinsic optical resonant cavity due to the unequal refractive indices in their metallic surfaces and bulk.(16, 63) The resonant cavity leads to enhancement of phase shifts and thus the holographic imaging. They calculated phase diagram of original images and print on $Sb_2Te_3$ thin film using direct laser writing methods, which is often used for 3D printing.(79-81) They achieved high quality holographic imaging on nanometric holograms. The work paves a way towards integrating holography with flat electronic devices for optical imaging, data storage and information security.

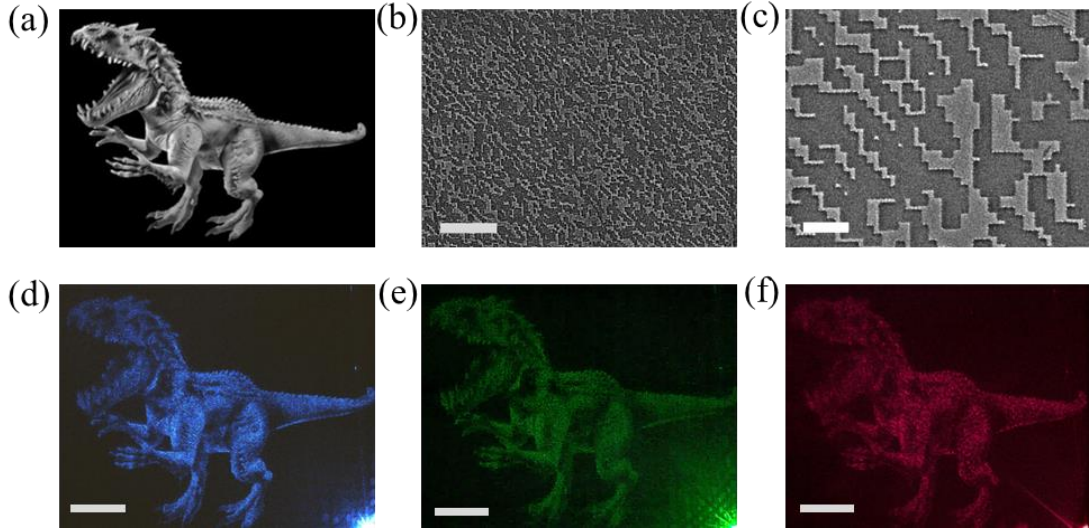

Figure 7. Nanometric $Sb_2Te_3$ thin film holograms and reconstructed images. (a) Original image of the dinosaur object. (b-c) SEM images of the laser printed hologram patterns. (d-f) Holographic images captured by illuminating the nanometric holograms using 445, 532 and 632 nm continuous wavelaser beams. Reprinted with permission from [Yue, 2017], Nature Communications, (2017) 15354. © 2017 Macmillan Publishers Limited, part of Springer Nature.

### 3.3 Ultrathin flat lens

An ultrathin double-focusing lens, capable of simultaneously generating a plasmonic focus in the near-field region and a Fresnel-zone-plate based diffraction-limited focal spot in the far-field region, was designed using $Sb_2Te_3$ thin films. The $Sb_2Te_3$ thin films hold high-index in the bulk and plasmonic excitations on the surface.(82) The double-focusing flat lens opens new opportunities for future compact devices with versatile functionalities, with potential applications ranging from optical imaging and information technology.

### 3.4 Near-infrared photodetector

Photodetectors are sensors of light, which can convert light photons into photocurrent or photovolatge.(83-86) Topological insulators have an energy gap in the bulk and a gapless surface state consisting of a single Dirac cone. Low-frequency optical absorption due to the surface state is universally determined by the fine-structure constant. When the thickness of these 3D topological insulators is reduced, they become quasi-two-dimensional insulators with enhanced absorbance. The two-dimensional insulators can be topologically trivial or nontrivial depending on the thickness, and we predict that the optical absorption is larger for topological nontrivial case compared with the trivial case. Since the surface state is intrinsically gapless, we propose its potential application in wide bandwidth, high-performance photodetection covering a broad spectrum ranging from terahertz to infrared. The performance of photodetection can be dramatically enhanced when the thickness is reduced to several quintuple layers with a widely tunable band gap depending on the thickness.(17)

Plank *et al.* demonstrated terahertz/infrared radiation induced photogalvanic effect in 3D TI $(Bi_{1-x}Sb_x)_2Te_3$.(87) They found that which the observed photogalvanic effect is sensitive to the surface symmetry and scattering details and can be applied to study the high frequency conductivity of the surface states. In particular, measuring the polarization dependence of the photogalvanic current and scanning with a micrometre sized beam spot across the sample, provides access to topographical inhomogeneities in the electronic properties of the surface states and the local domain orientation.

Zheng, *et al.* fabricated a near infrared (NIR) light photodetector based on a $Sb_2Te_3$ thin film, which was grown on sapphire by molecular beam epitaxy (MBE). Electrical analysis reveals that the resistance of the TIdecreases with increasing temperature in the temperature range of 8.5–300 K. Further optoelectronic characterization showed that the as-fabricated photodetector exhibits obvious sensitivity to 980 nm light illumination. The responsivity, photoconductive gain and detectivity were estimated to be 21.7 A/W, 27.4 and $1.22 \times 10^{11}$ Jones, respectively, which are much better than those of other topological insulators based devices. This study suggests that the present NIR photodetector may have potential application in future optoelectronic devices.(88)

Sharma *et al.* reported a high performance broadband photodetector based on $Bi_2Se_3$ nanowires. (89) They fabricated the $Bi_2Se_3$ nanowires using focused ion beam (FIB) and found they can be used for ultrasensitive visible-NIR photodetectors. They observed efficient electron hole pair generation in the $Bi_2Se_3$ nanowire under the illumination of visible (532 nm) and IR light (1064 nm). They observed photo-responsivity of up to 300 A/W.

### 3.5 Saturable absorber

Yu, *et al.* demonstrates that topological insulator can modulate the operation of a bulk solid-state laser by taking advantage of its saturable absorption. Their result suggests that topological insulators are potentially attractive as broadband pulsed modulators for the generation of short and ultrashort pulses in bulk solid-state lasers, in addition to other promising applications in physics and computing.(90)

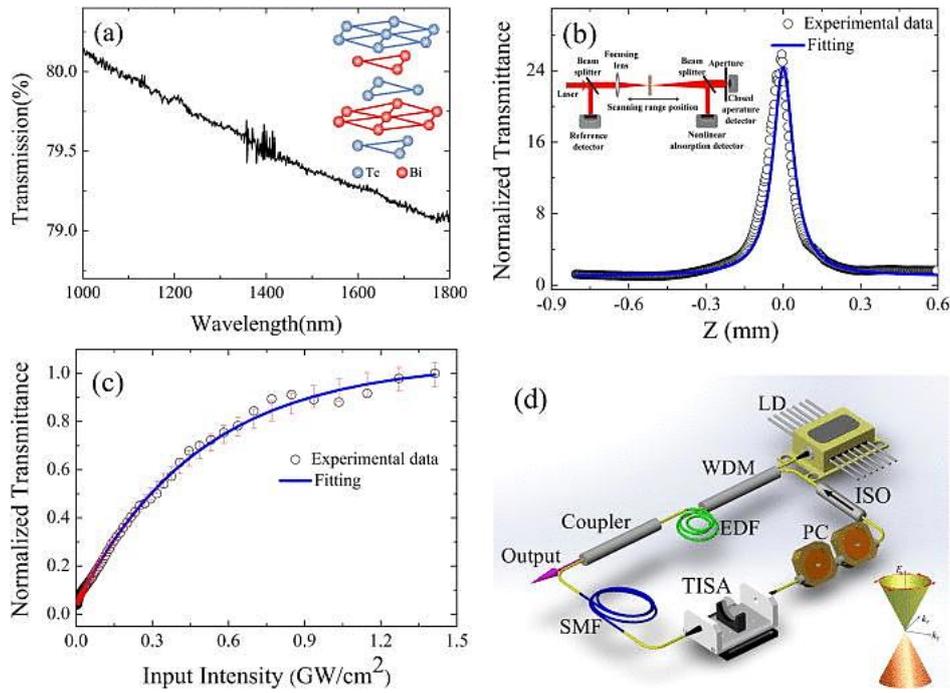

Figure 8. Ultra-short pulse generation by topological insulator material $Bi_2Te_3$ that works as a saturable absorber. (a) The near infrared linear absorption spectra of the $Bi_2Te_3$. (b) A typical Z-scan peak curve of $Bi_2Te_3$ at 1550 nm. (c) The corresponding nonlinear saturable absorption curve. (d) Schematic of the fiber laser. Reprinted with permission from [Yue, 2017]. (91).

Zhao, *et al.* showed that the topological insulator material $Bi_2Te_3$ is a saturable absorber with high modulation depth at 1.55 μm. The $Bi_2Te_3$ based saturable absorber device was fabricated and used as a passive mode locker for ultrafast pulse formation at the telecommunication band.(91) 3-μm mid-infrared pulse was also generated using the $Bi_2Te_3$ as the saturable absorber.(92) The $Bi_2Te_3$ shows a low saturable peak intensity of 2.12 $MW/cm^2$ and a high modulation depth of 51.3%. Lee, *et al.* found that the bulk-structured $Bi_2Te_3$ layer can provide sufficient nonlinear saturable absorption for femtosecond mode-locking.(93) They used the $Bi_2Te_3$ as an ultrafast mode-locker to generate femtosecond pulses from an all-fiberized cavity. They presented that stable soliton pulses with a temporal width of ~600 femtosecond can be produced at 1.55 μm from an erbium fiber ring cavity.

Large energy, wavelength tunable Q-switched erbium-doped fiber laser was fabricated using the $Bi_2Te_3$.(94) The saturating intensity was ~57 $MW/cm^2$ and the modulation depth was ~22%. The single pulse energy is ~1.5 μJ and the saturable absorption operation from ~1.51 μm to ~1.58 μm. 2 GHz passively harmonic mode-locked fiber laser was also achieved by a microfiber-based $Bi_2Te_3$ saturable absorber.(95) The fiber laser could operate at the pulse repetition rate of 2.04 GHz under a pump power of 126 mW.

$Bi_2Te_3$ nanoparticles have shown the broadband saturable absorption at 0.8 μm and 1.57 μm. They were employed as nonlinear saturable absorbers to passively mode-lock the erbium-doped fiber lasers for sub-400 fs pulse generations.(96) Xu, *et al.* reported that the $Bi_2Te_3$ exhibited a nonlinear absorption response.(97) They showed that the $Bi_2Te_3$ sheets have saturation absorption intensity of 1.1 W/cm$^2$ at 1.0 μm. A Q-switching pulsed laser was made in a 1.0 μm Nd:YVO$_4$ laser where the threshold absorbed pump power was 31 mW. A pulse duration of 97 ns was observed with an average power of 26.1 mW. A Q-switched laser at 1.3 μm was also realized with a pulse duration of 93 ns.

Wavelength-tunable picosecond soliton fiber laser was achieved using topological insulator material $Bi_2Se_3$ as a mode locker.(98) An optical pulse with ~660 fs was generated at wavelength of 1.55 μm. A modulation depth of 98% and a saturation intensity of 0.49 GW/cm$^2$ were observed. Femtosecond pulse was also generated from $Bi_2Se_3$ mode-locked fiber laser.(99) The used $Bi_2Se_3$ saturable absorber has a low saturable optical intensity of 12 MW/cm$^2$ and a modulation depth of ~3.9%. The mode-locking operation was realized at 25 mW.

Dou, *et al.* demonstrated a mode-locked ytterbium-doped fiber laser based on the $Bi_2Se_3$.(100) The measured modulation depth of the $Bi_2Se_3$ film was 5.2%. When the $Bi_2Se_3$ film was used in the Yb-doped fiber laser, the mode locked pulses have pulse energy of 0.756 nJ, pulse width of 46 ps and the repetition rate of 44.6 MHz. The maximum average output power was 33.7 mW. When the pump power exceeded 270 mW, the laser can operate in multiple pulse state that six-pulse regime can be realized.

Yu, *et al.* showed a high-repetition-rate Q-switched fiber laser using the $Bi_2Se_3$ film.(101) The $Bi_2Se_3$ film had a low saturable optical intensity of 11 MW/cm$^2$. By inserting the absorber film into an Erbium-doped fiber laser, a high-repetition Q-switched laser with the repetition rates from 459 kHz to 940 kHz was made. The maximum output power was 22.35 mW with the shortest pulse duration of 1.9 μs.

Luo, *et al.* demonstrated that a 1.06μm Q-switched ytterbium-doped fiber laser using few-layer $Bi_2Se_3$ as a saturable absorber.(101) The few-layer $Bi_2Se_3$ has a low saturable optical intensity of 53 MW/cm$^2$. By inserting $Bi_2Se_3$ into the YDF laser cavity, stable Q-switching operation at 1.06 μm is achieved. The Q-switched pulses have the pulse duration of 1.95 μs, the pulse energy of 17.9 nJ and a tunable pulse-repetition-rate from 8.3 to 29.1 kHz.

*Gao, et al.* reported a *Q*-switched mode-locked erbium-doped fiber laser based on the $Bi_2Se_3$ deposited fiber taper.(102) Due to the low saturation intensity, stable Q-switched mode-locked fiber lasers centered at 1.56 μm can be generated at a pump power of 10 mW.

These results exhibit that the topological insulator materials $Bi_2Te_3$ and $Bi_2Se_3$ are promising optical materials for constructing broadband, miniature and integrated high-energy pulsed laser systems with low power consumption.

## References

1. Hasan MZ & Kane CL (2010) Colloquium: Topological insulators. *Reviews of Modern Physics* 82(4):3045-3067.
2. Qi X-L & Zhang S-C (2011) Topological insulators and superconductors. *Reviews of Modern Physics* 83(4):1057-1110.
3. Yue Z (2013) Electronic transport, magnetic and thermal properties of gapless materials. *UoW PhD Thesis* 4138.
4. Bernevig BA, Hughes TL, & Zhang S-C (2006) Quantum Spin Hall Effect and Topological Phase Transition in HgTe Quantum Wells. *Science* 314(5806):1757-1761.
5. Hsieh D*, et al.* (2008) A topological Dirac insulator in a quantum spin Hall phase. *Nature* 452(7190):970-974.
6. Moore JE (2010) The birth of topological insulators. *Nature* 464(7286):194-198.
7. Wang X-L, Dou SX, & Zhang C (2010) Zero-gap materials for future spintronics, electronics and optics. *NPG Asia Mater* 2:31-38.
8. Ando Y (Topological Insulator Materials. *Journal of the Physical Society of Japan* 82(Copyright (C) 2013 The Physical Society of Japan):102001.
9. Chen YL*, et al.* (2009) Experimental Realization of a Three-Dimensional Topological Insulator, Bi2Te3. *Science* 325(5937):178-181.
10. Yue Z, Chen Q, & Wang X (2016) Thermoelectric signals of state transition in polycrystalline SmB6. *Journal of Physics: Condensed Matter* 28(35):355801.
11. Zhang H*, et al.* (2009) Topological insulators in Bi2Se3, Bi2Te3 and Sb2Te3 with a single Dirac cone on the surface. *Nat Phys* 5(6):438-442.
12. Ando Y (2013) Topological Insulator Materials. *Journal of the Physical Society of Japan* 82(10):102001.
13. Di Pietro P*, et al.* (2013) Observation of Dirac plasmons in a topological insulator. *Nat Nano* 8(8):556-560.
14. Ou J-Y*, et al.* (2014) Ultraviolet and visible range plasmonics in the topological insulator Bi1.5Sb0.5Te1.8Se1.2. *Nat Commun* 5.
15. Yue Z, Cai B, Wang L, Wang X, & Gu M (2016) Intrinsically core-shell plasmonic dielectric nanostructures with ultrahigh refractive index. *Science Advances* 2(3).
16. Yue Z, Xue G, Liu J, Wang Y, & Gu M (2017) Nanometric holograms based on a topological insulator material. 8:15354.
17. Zhang X, Wang J, & Zhang S-C (2010) Topological insulators for high-performance terahertz to infrared applications. *Physical Review B* 82(24):245107.
18. Kane CL & Mele EJ (2005) Quantum Spin Hall Effect in Graphene. *Physical Review Letters* 95(22):226801.
19. Bernevig BA & Zhang S-C (2006) Quantum Spin Hall Effect. *Physical Review Letters* 96(10):106802.
20. König M*, et al.* (2007) Quantum Spin Hall Insulator State in HgTe Quantum Wells. *Science* 318(5851):766-770.
21. Roth A*, et al.* (2009) Nonlocal Transport in the Quantum Spin Hall State. *Science* 325(5938):294-297.

22. Qi X-L, Hughes TL, & Zhang S-C (2008) Topological field theory of time-reversal invariant insulators. *Physical Review B* 78(19):195424.
23. Qi X-L, Li R, Zang J, & Zhang S-C (2009) Inducing a Magnetic Monopole with Topological Surface States. *Science* 323(5918):1184-1187.
24. Qi X-L, Hughes TL, & Zhang S-C (2010) Chiral topological superconductor from the quantum Hall state. *Physical Review B* 82(18):184516.
25. Fu L & Kane CL (2008) Superconducting Proximity Effect and Majorana Fermions at the Surface of a Topological Insulator. *Physical Review Letters* 100(9):096407.
26. Mourik V*, et al.* (2012) Signatures of Majorana Fermions in Hybrid Superconductor-Semiconductor Nanowire Devices. *Science* 336(6084):1003-1007.
27. Das A*, et al.* (2012) Zero-bias peaks and splitting in an Al-InAs nanowire topological superconductor as a signature of Majorana fermions. *Nat Phys* 8(12):887-895.
28. Rokhinson LP, Liu X, & Furdyna JK (2012) The fractional a.c. Josephson effect in a semiconductor-superconductor nanowire as a signature of Majorana particles. *Nat Phys* 8(11):795-799.
29. Stern A & Lindner NH (2013) Topological Quantum Computation—From Basic Concepts to First Experiments. *Science* 339(6124):1179-1184.
30. He QL*, et al.* (2017) Chiral Majorana fermion modes in a quantum anomalous Hall insulator–superconductor structure. *Science* 357(6348):294-299.
31. Yu R*, et al.* (2010) Quantized Anomalous Hall Effect in Magnetic Topological Insulators. *Science* 329(5987):61-64.
32. Chang C-Z*, et al.* (2013) Experimental Observation of the Quantum Anomalous Hall Effect in a Magnetic Topological Insulator. *Science* 340(6129):167-170.
33. Yue Z*, et al.* (2015) Crossover of Magnetoresistance from Fourfold to Twofold Symmetry in SmB6 Single Crystal, a Topological Kondo Insulator. *Journal of the Physical Society of Japan* 84(4):044717.
34. Yue Z, Zhu C, Dou S, & Wang X (2012) Observation of field-induced polarization of valleys in p-type Sb$_2$Te$_3$ single crystals. *Physical Review B* 86(19):195120.
35. Yue ZJ, Wang XL, & Dou SX (2012) Angular-dependences of giant in-plane and interlayer magnetoresistances in Bi2Te3 bulk single crystals. *Applied Physics Letters* 101(15):152107.
36. Yue Z, Wang X, & Dou S (2012) Giant interlayer magnetoresistances and strong anisotropy in p-type Sb2Te3 single crystals. *Integrated Ferroelectrics* 140(1):155-160.
37. Yue Z, Wang X, & Yan S-S (2015) Semimetal-semiconductor transition and giant linear magnetoresistances in three-dimensional Dirac semimetal Bi0. 96Sb0. 04 single crystals. *Applied Physics Letters* 107(11):112101.
38. Xiang F-X, Wang X-L, Veldhorst M, Dou S-X, & Fuhrer MS (2015) Observation of topological transition of Fermi surface from a spindle torus to a torus in bulk Rashba spin-split BiTeCl. *Physical Review B* 92(3):035123.
39. Yue Z*, et al.* (2012) Giant and anisotropic magnetoresistances in p-type Bi-doped Sb2Te3 bulk single crystals. *EPL (Europhysics Letters)* 100(1):17014.
40. Xia B*, et al.* (2013) Indications of surface-dominated transport in single crystalline nanoflake devices of topological insulator Bi$_{1.5}$Sb$_{0.5}$Te$_{1.8}$Se$_{1.2}$. *Physical Review B* 87(8):085442.
41. Sulaev A*, et al.* (2015) Electrically Tunable In-Plane Anisotropic Magnetoresistance in Topological Insulator BiSbTeSe2 Nanodevices. *Nano Letters* 15(3):2061-2066.
42. Bao L*, et al.* (2012) Weak Anti-localization and Quantum Oscillations of Surface States in Topological Insulator Bi2Se2Te. *Sci. Rep.* 2.
43. Qu D-X, Hor YS, Xiong J, Cava RJ, & Ong NP (2010) Quantum Oscillations and Hall Anomaly of Surface States in the Topological Insulator Bi2Te3. *Science* 329(5993):821-824.
44. Taskin AA & Ando Y (2009) Quantum oscillations in a topological insulator Bi$_{1-x}$Sb$_x$. *Physical Review B* 80(8):085303.

45. He H, *et al.* (2012) High-field linear magneto-resistance in topological insulator Bi[sub 2]Se[sub 3] thin films. *Applied Physics Letters* 100(3):032105.
46. Tang H, Liang D, Qiu RLJ, & Gao XPA (2011) Two-Dimensional Transport-Induced Linear Magneto-Resistance in Topological Insulator Bi2Se3 Nanoribbons. *ACS Nano* 5(9):7510-7516.
47. Chiu S-P & Lin J-J (2013) Weak antilocalization in topological insulator Bi_{2}Te_{3} microflakes. *Physical Review B* 87(3):035122.
48. Wang X, Du Y, Dou S, & Zhang C (2012) Room Temperature Giant and Linear Magnetoresistance in Topological Insulator Bi_{2}Te_{3} Nanosheets. *Physical Review Letters* 108(26):266806.
49. Abrikosov AA (1998) Quantum magnetoresistance. *Physical Review B* 58(5):2788-2794.
50. Abrikosov AA (1999) Quantum magnetoresistance of layered semimetals. *Physical Review B* 60(6):4231-4234.
51. Lu H-Z, Shi J, & Shen S-Q (2011) Competition between Weak Localization and Antilocalization in Topological Surface States. *Physical Review Letters* 107(7):076801.
52. Chen J, *et al.* (2010) Gate-Voltage Control of Chemical Potential and Weak Antilocalization in Bi_{2}Se_{3}. *Physical Review Letters* 105(17):176602.
53. He H-T, *et al.* (2011) Impurity Effect on Weak Antilocalization in the Topological Insulator Bi_{2}Te_{3}. *Physical Review Letters* 106(16):166805.
54. Liu M, *et al.* (2012) Crossover between Weak Antilocalization and Weak Localization in a Magnetically Doped Topological Insulator. *Physical Review Letters* 108(3):036805.
55. Lang M, *et al.* (2012) Competing Weak Localization and Weak Antilocalization in Ultrathin Topological Insulators. *Nano Letters* 13(1):48-53.
56. Lu H-Z & Shen S-Q (2011) Weak localization of bulk channels in topological insulator thin films. *Physical Review B* 84(12):125138.
57. Xiong J, *et al.* (2012) High-field Shubnikov–de Haas oscillations in the topological insulator Bi_{2}Te_{2}Se. *Physical Review B* 86(4):045314.
58. Chang LL, Sakaki H, Chang CA, & Esaki L (1977) Shubnikov—de Haas Oscillations in a Semiconductor Superlattice. *Physical Review Letters* 38(25):1489-1493.
59. Ishiwata S, *et al.* (2013) Extremely high electron mobility in a phonon-glass semimetal. *Nat Mater* 12(6):512-517.
60. Wang W, *et al.* (2013) Large Linear Magnetoresistance and Shubnikov-de Hass Oscillations in Single Crystals of YPdBi Heusler Topological Insulators. *Sci. Rep.* 3.
61. Tian M, *et al.* (2013) Dual evidence of surface Dirac states in thin cylindrical topological insulator Bi2Te3 nanowires. *Sci. Rep.* 3.
62. Yue Z, Xue G, & Gu M (2016) Resonant cavity-enhanced holographic imaging in ultrathin topological insulator films. *Frontiers in Optics 2016*, (Optical Society of America), p FTu3F.1.
63. Yue Z, Xue G, Liu J, Wang Y, & Gu M (2017) Nanometric holograms based on a topological insulator material. *Nature Communications* 8:15354.
64. Yue Z & Gu M (2016) Resonant cavity-enhanced holographic imaging in ultrathin topological insulator films. *Frontiers in Optics*, (Optical Society of America), p FTu3F. 1.
65. Peng H, *et al.* (2012) Topological insulator nanostructures for near-infrared transparent flexible electrodes. *Nat Chem* 4(4):281-286.
66. Ge L, Zhan T, Han D, Liu X, & Zi J (2014) Unusual electromagnetic scattering by cylinders of topological insulator. *Opt. Express* 22(25):30833-30842.
67. Maier SA (2007) *Plasmonics: fundamentals and applications* (Springer Science & Business Media).
68. Fang Z & Zhu X (2013) Plasmonics in nanostructures. *Advanced Materials* 25(28):3840-3856.
69. Zhang Q, *et al.* (2014) Graphene surface plasmons at the near-infrared optical regime. 4:6559.
70. Tan C, *et al.* (2018) Nanograting-assisted generation of surface plasmon polaritons in Weyl semimetal WTe2. *Optical Materials* 86:421-423.

71. Zhao M, *et al.* (2016) Actively Tunable Visible Surface Plasmons in Bi2Te3 and their Energy-Harvesting Applications. *Advanced Materials* 28(16):3138-3144.
72. Zhao M, *et al.* (2015) Visible Surface Plasmon Modes in Single Bi2Te3 Nanoplate. *Nano Letters* 15(12):8331-8335.
73. Yuan J, *et al.* (2017) Infrared Nanoimaging Reveals the Surface Metallic Plasmons in Topological Insulator. *ACS Photonics*.
74. Yue Z, Ren H, Wei S, Lin J, & Gu M (2018) Angular-momentum nanometrology in an ultrathin plasmonic topological insulator film. *Nature Communications* 9(1):4413.
75. McIver JW, Hsieh D, Steinberg H, Jarillo Herrero P, & Gedik N (2012) Control over topological insulator photocurrents with light polarization. *Nat Nano* 7(2):96-100.
76. Chen S, *et al.* (2014) Broadband optical and microwave nonlinear response in topological insulator. *Opt. Mater. Express* 4(4):587-596.
77. Giorgianni F, *et al.* (2016) Strong nonlinear terahertz response induced by Dirac surface states in Bi2Se3 topological insulator. 7:11421.
78. Lu S, *et al.* (2013) Third order nonlinear optical property of Bi2Se3. *Opt. Express* 21(2):2072-2082.
79. Goi E, Yue Z, Cumming BP, & Gu M (2018) A Layered-Composite Nanometric Sb2Te3 Material for Chiral Photonic Bandgap Engineering. *physica status solidi (a)* 215(14):1800152.
80. Goi E, Yue ZJ, Cumming BP, & Gu M (2017) Two-photon fabrication of type I optical Weyl points in photonic crystals. *Frontiers in Optics 2017*, (Optical Society of America), p FW6C.5.
81. Yu H, Zhang Q, Yue Z, & Gu M (2017) Three-dimensional Direct Laser Writing of Neuron-inspired Structures. *Frontiers in Optics 2017*, (Optical Society of America), p FTu5D.2.
82. Goi E, Yue ZJ, Cumming BP, & Gu M (2016) Complete bandgap in three-dimensional chiral gyroid photonic crystals for topological photonics. *Lasers and Electro-Optics (CLEO), 2016 Conference on*, (IEEE), pp 1-2.
83. Yue ZJ, *et al.* (2009) Thickness-dependent photovoltaic effects in miscut Nb-doped SrTiO3 single crystals. *Journal of Physics D: Applied Physics* 43(1):015104.
84. Li X, *et al.* (2010) Voltage tunable photodetecting properties of La 0.4 Ca 0.6 MnO 3 films grown on miscut LaSrAlO 4 substrates. *Applied Physics Letters* 97(4):044104.
85. Yue Z, *et al.* (2011) Photo-induced magnetoresistance enhancement in manganite heterojunction at room temperature. *Journal of Physics D: Applied Physics* 44(9):095103.
86. Ni H, *et al.* (2012) Magnetical and electrical tuning of transient photovoltaic effects in manganite-based heterojunctions. *Opt. Express* 20(103):A406-A411.
87. Plank H, *et al.* (2016) Opto-electronic characterization of three dimensional topological insulators. *Journal of Applied Physics* 120(16):165301.
88. Zheng K, *et al.* (2015) Optoelectronic characteristics of a near infrared light photodetector based on a topological insulator Sb2Te3 film. *Journal of Materials Chemistry C* 3(35):9154-9160.
89. Sharma A, Bhattacharyya B, Srivastava AK, Senguttuvan TD, & Husale S (2016) High performance broadband photodetector using fabricated nanowires of bismuth selenide. 6:19138.
90. Yu H, *et al.* (2013) Topological insulator as an optical modulator for pulsed solid-state lasers. *Laser & Photonics Reviews* 7(6):L77-L83.
91. Zhao C, *et al.* (2012) Ultra-short pulse generation by a topological insulator based saturable absorber. *Applied Physics Letters* 101(21):211106.
92. Li J, *et al.* (2015) 3-μm mid-infrared pulse generation using topological insulator as the saturable absorber. *Opt. Lett.* 40(15):3659-3662.
93. Lee J, Koo J, Jhon YM, & Lee JH (2014) A femtosecond pulse erbium fiber laser incorporating a saturable absorber based on bulk-structured Bi2Te3 topological insulator. *Opt. Express* 22(5):6165-6173.

94. Chen Y, *et al.* (2014) Large Energy, Wavelength Widely Tunable, Topological Insulator Q-Switched Erbium-Doped Fiber Laser. *IEEE Journal of Selected Topics in Quantum Electronics* 20(5):315-322.
95. Luo Z-C, *et al.* (2013) 2 GHz passively harmonic mode-locked fiber laser by a microfiber-based topological insulator saturable absorber. *Opt. Lett.* 38(24):5212-5215.
96. Lin Y-H, *et al.* (2015) Using n- and p-Type Bi2Te3 Topological Insulator Nanoparticles To Enable Controlled Femtosecond Mode-Locking of Fiber Lasers. *ACS Photonics* 2(4):481-490.
97. Xu J-L, *et al.* (2015) Ultrasensitive nonlinear absorption response of large-size topological insulator and application in low-threshold bulk pulsed lasers. 5:14856.
98. Zhao C, *et al.* (2012) Wavelength-tunable picosecond soliton fiber laser with Topological Insulator: Bi2Se3 as a mode locker. *Opt. Express* 20(25):27888-27895.
99. Liu H, *et al.* (2014) Femtosecond pulse generation from a topological insulator mode-locked fiber laser. *Opt. Express* 22(6):6868-6873.
100. Dou Z, *et al.* (2014) Mode-locked ytterbium-doped fiber laser based on topological insulator: Bi2Se3. *Opt. Express* 22(20):24055-24061.
101. Yu Z, *et al.* (2014) High-repetition-rate Q-switched fiber laser with high quality topological insulator Bi2Se3 film. *Opt. Express* 22(10):11508-11515.
102. Gao L, *et al.* (2014) Q-switched mode-locked erbium-doped fiber laser based on topological insulator Bi2Se3 deposited fiber taper. *Applied Optics* 53(23):5117-5122.